\documentclass[twocolumn]{aastex62}
\usepackage{enumitem}
\usepackage{courier}
\graphicspath{{./}{figures/}}
\usepackage[T1]{fontenc}

\begin{document}
\title{Rotation Distributions around the Kraft Break with TESS and \textit{Kepler}: The Influences of Age, Metallicity, and Binarity}

\received{26 August 2021}
\revised{21 March 2022}
\accepted{22 March 2022}

\author[0000-0003-1719-5046]{Ellis A. Avallone}
\affiliation{Institute for Astronomy, University of Hawai‘i at Mānoa, 2680 Woodlawn Drive, Honolulu, HI 96822, USA}

\author[0000-0002-4818-7885]{Jamie N. Tayar}
\altaffiliation{NASA Hubble Fellow}
\affiliation{Institute for Astronomy, University of Hawai‘i at Mānoa, 2680 Woodlawn Drive, Honolulu, HI 96822, USA}
\affiliation{Department of Astronomy, University of Florida, Bryant Space Science Center, Stadium Road, Gainesville, FL 32611, USA }

\author[0000-0002-4284-8638]{Jennifer L. van Saders}
\affiliation{Institute for Astronomy, University of Hawai‘i at Mānoa, 2680 Woodlawn Drive, Honolulu, HI 96822, USA}

\author[0000-0002-2580-3614]{Travis A. Berger}
\altaffiliation{NASA Postdoctoral Program Fellow}
\affiliation{Exoplanets and Stellar Astrophysics Laboratory, Code 667, NASA Goddard Space Flight Center, Greenbelt, MD, 20771, USA}
\affiliation{Institute for Astronomy, University of Hawai‘i at Mānoa, 2680 Woodlawn Drive, Honolulu, HI 96822, USA}

\author[0000-0002-9879-3904]{Zachary R. Claytor}
\affiliation{Institute for Astronomy, University of Hawai‘i at Mānoa, 2680 Woodlawn Drive, Honolulu, HI 96822, USA}

\author[0000-0002-1691-8217]{Rachael~L.~Beaton}
\altaffiliation{Much of this work was completed while this author was a NASA Hubble Fellow at Princeton University.}
\affiliation{Carnegie-Princeton Fellow}
\affiliation{Department of Astrophysical Sciences, Princeton University, 4 Ivy Lane, Princeton, NJ~08544}
\affiliation{The Observatories of the Carnegie Institution for Science, 813 Santa Barbara St., Pasadena, CA~91101}

\author{Johanna Teske}
\affiliation{Earth and Planets Laboratory, Carnegie Institution for Science, 5241 Broad Branch Road, NW, Washington, DC 20015, USA}

\author[0000-0003-4556-1277]{Diego Godoy-Rivera}
\affiliation{Department of Astronomy, The Ohio State University, 140 West 18th Avenue, Columbus, OH 43210, USA}

\author{Kaike Pan}
\affiliation{Apache Point Observatory and New Mexico State
University, P.O. Box 59, Sunspot, NM, 88349-0059, USA}

\begin{abstract}
   Stellar rotation is a complex function of mass, metallicity, and age and can be altered by binarity. To understand the importance of these parameters in main sequence stars, we have assembled a sample of observations that spans a range of these parameters using a combination of observations from The Transiting Exoplanet Survey Satellite (TESS) and the \textit{Kepler} Space Telescope. We find that while we can measure rotation periods and identify other classes of stellar variability (e.g., pulsations) from TESS lightcurves, instrument systematics prevent the detection of rotation signals longer than the TESS orbital period of 13.7 days. Due to this detection limit, we also utilize rotation periods constrained using rotational velocities measured by the APOGEE spectroscopic survey and radii estimated using the \textit{Gaia} mission for both TESS and \textit{Kepler} stars. From these rotation periods, we 1) find we can track rotational evolution along discrete mass tracks as a function of stellar age, 2) find we are unable to recover trends between rotation and metallicity that were observed by previous studies, and 3) note that our sample reveals that wide binary companions do not affect rotation, while close binary companions cause stars to exhibit more rapid rotation than single stars.
   
\end{abstract}

\section{Introduction}\label{intro}
Stars in nature rotate, and rotation is an important ingredient in stellar evolution for the most rapid rotators. Stellar models, however, often do not take rotation into account or simplify rotation when they do \citep[][]{dantona_1984,maeder_1989A&A...210..155M,Pinsonneault_1989,ekstrom_2012,2013ApJ...776...67V,choi_mist_2016,choi_2017,Ostrowski_2017}. This is primarily because the physics governing angular momentum transport and loss are complex and uncertain.

Angular momentum loss occurs as a consequence of magnetized stellar winds, which carry angular momentum away from stars and cause stellar spin-down as stars evolve \citep[][]{Weber_davis_1967,Skumanich_1972}. Most braking laws predict angular momentum loss that goes as $dJ/dt \propto \omega ^3$, where the angular momentum loss strongly depends on the rotational velocity of the star \citep[][]{1988ApJ...333..236K}. This has the effect of producing tight rotational sequences in intermediate-age open clusters, despite the fact that stars begin their lives with a wide range of rotation periods \citep[e.g.][]{Meibom_2011,Meibom_2015,Barnes_2016,Douglas_2017,Rebull_2017,Curtis_2020,Fritzewski_2020,godoy_rivera_2021}. 

\pagebreak

The primary focus of previous observational studies of rotation as it relates to stellar evolution has been on main sequence stars that experience substantial spin-down, with some focus on cool evolved stars \citep[e.g.][]{mamajek_2008,Meibom_2009,Meibom_2011,nielsen_2013,2013ApJ...776...67V,2018ApJ...868..150T}. The \textit{Kepler} Space Telescope enabled many of these studies of rotation across the main sequence \citep[e.g ][]{mcquillan_2013MNRAS.432.1203M,mcquillan_2013ApJ...775L..11M_2,2014ApJS..211...24M,Amard_2020,Simonian_2020}. Its successor, the Transiting Exoplanet Survey Satellite \citep[TESS;][]{2015JATIS...1a4003R}, has enabled further studies of rotation across stellar populations \citep[e.g.][]{Lu_astrea_2020,Martins_2020_tessrot}. 

As with \textit{Kepler}, rotation studies with TESS rely on measurements from starspot modulation in lightcurves. These rotation rates are often preferred to rotation rates measured via other methods due to their independence from stellar inclination. Previous work has shown that rotation periods from starspot modulation agree with rotation periods derived from spectroscopic $v\sin i$ \citep{nielsen_2013,Simonian_2020} and asteroseismically inferred rotation periods \citep[][]{reinhold_2015,beck_2017A&A...602A..63B,hall_2021}. 

Attempts at studying rotation with TESS lightcurves, however, have been met with several challenges \citep[e.g.][]{Lu_astrea_2020,Martins_2020_tessrot}. The $27.4$-day observing sectors in TESS limit the measurement of rotation signals beyond tens of days for much of the survey area. Stars observed in the southern continuous viewing zone (SCVZ), however, have a combined observing baseline of $\sim$ 2 years from the 13 southern hemisphere observing sectors and extended mission observations. Although the combined baseline of these observations should enable studies of rotation signals that span multiple TESS observing sectors, attempts at measuring rotation periods longer than the length of a single sector ($27.4$ days) have been unsuccessful with standard rotation period measurement methods \citep[e.g.][]{Guenther_2020_tessrot,Martins_2020_tessrot}, although there has been some success with machine learning-based methods of extracting rotation signals \citep[e.g. with neural networks][]{Lu_astrea_2020,Claytor_2021}. These prior studies using traditional methods of measuring rotation periods from lightcurves have suggested that challenges in measuring rotation signals across multiple sectors in TESS are likely due to scattered light at the start and end of observing sectors and signals from the orbital motion of the spacecraft around Earth. These timescale constraints have limited the types of stars where we can effectively probe rotation. 

We therefore focus our efforts here on main sequence stars near the Kraft Break, which tend to be rapid rotators. The Kraft Break is a transition that is seen across stellar effective temperatures ($T_\mathrm{eff}$). Hot stars ($T_\mathrm{eff}>6250$ K) typically experience little to no angular momentum loss due to the presence of vanishingly thin convective zones that cannot sufficiently support a magnetized wind. Cool stars ($T_\mathrm{eff}<6250$ K) typically experience substantial angular momentum loss due to the presence of large convection zones that can sustain a magnetized wind \citep[][]{1967ApJ...150..551K}.  Main sequence stars above the Kraft Break were the first stars to have reliable rotation measurements and lend themselves well to spectroscopic investigations due to their rapid rotation \citep[][]{1967ApJ...150..551K}. 

Our focus on stars near the Kraft Break can provide a look into the distribution of rotation periods at the transition between stars which experience Solar-like spin-down and stars which experience little to no angular momentum loss. Some work has already been done to characterize the main sequence distribution of rotation periods as a function of other stellar parameters with the \textit{Kepler} Space Telescope \citep[e.g ][]{2014ApJS..211...24M,Amard_2020,Simonian_2020}.

Here, we measure rotation periods from starspot modulation for a sample of stars observed by TESS and augment this sample with stars whose rotation periods were measured from \textit{Kepler} lightcurves by \citet{nielsen_2013}, \citet{reinhold_2013}, and \citet{2014ApJS..211...24M}. We use spectroscopic parameters measured by the Apache Point Observatory Galactic Evolution Experiment \citep[APOGEE;][]{2017AJ....154...94M,APOGEE_DR16,Jonsson_2020} and photometric radii estimated using the \textit{Gaia} mission to carefully characterize each star in our sample and derive alternate rotation diagnostics from spectroscopic $v\sin i$ and \textit{Gaia} radii. The combination of these data enables us to obtain a more complete picture of rotation in a sample with significant detection bias in spots. 

\section{Data \& Methods}
\subsection{Observational Data}\label{data_obs}

We use Cycle 1 data from the TESS \citep[]{2015JATIS...1a4003R} SCVZ and spectroscopic parameters from an APOGEE \citep{2017AJ....154...94M,APOGEE_DR16,Jonsson_2020} External Program through Carnegie Observatories. The 37,000 stars in the TESS-APOGEE SCVZ sample have been observed near the ecliptic poles, where stars could be observed in a maximum of 13 TESS sectors, although the number of sectors with observations depends on the exact location of the star relative to the observing CCDs. These stars additionally have homogeneous spectroscopic characterization, making this the ideal sample to search for rotation periods.

\begin{figure}[ht!]
    \centering
    \includegraphics[width = 0.48\textwidth]{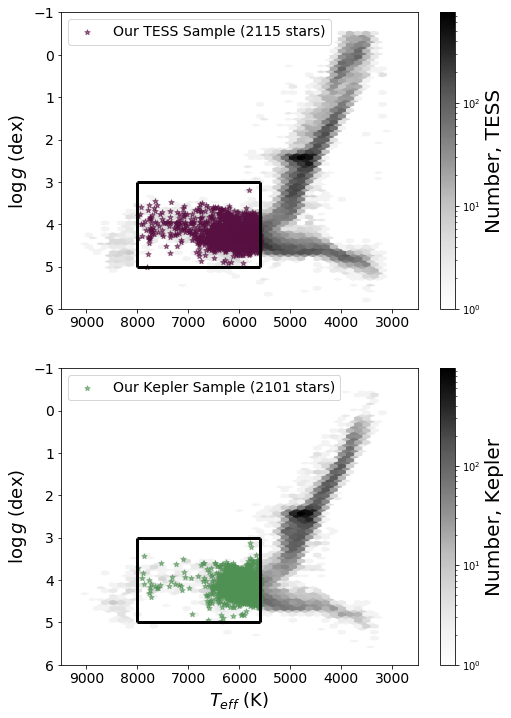}
    \caption{Kiel diagrams showing our selected sample of stars near the Kraft Break observed by both APOGEE and TESS (top) and by both APOGEE and \textit{Kepler} (bottom). The full APOGEE-TESS SCVZ (top) and APOGEE-\textit{Kepler} (bottom) samples are shown in the background, and our selected sample in both TESS (purple) and \textit{Kepler} (green) are shown.}
    \label{apo_sample}
\end{figure}

To select our sample, we constrain the APOGEE-TESS SCVZ sample to stars with $\log g$ values between $3$ and $5$ dex and $T_\mathrm{eff}$ values between $5600$ K and $8000$ K. These constraints on $\log g$ and $T_\mathrm{eff}$ allow us to encompass stars on the main sequence without risking contamination from giant-branch stars or stars with masses far below 1$M_{\bigodot}$. These constraints additionally exclude M type stars, where APOGEE $\log g$ values have high uncertainties \citep[][]{APOGEE_DR16,sarmento_2021A&A...649A.147S}. We restrict our sample to only include stars that have been observed with TESS's short cadence observing mode, as these stars have available lightcurves generated with the TESS Science Processing Operations Center pipeline \citep[SPOC;][]{Jenkins_2016}. We finally only select stars that have been observed across at least 9 TESS observing sectors to optimize our chance of detecting rotation signals that span multiple sectors. Of the 37,000 stars observed in the APOGEE-TESS SCVZ sample, our final TESS sample contains 2115 stars. The top panel in Figure \ref{apo_sample} shows a Kiel diagram of all stars in the APOGEE-TESS SCVZ sample along with our final TESS sample.

APOGEE data are collected with the $2.5$ meter Sloan Digital Sky Survey (SDSS) telescope \citep{SDSS_gunn2006} at Apache Point Observatory and the $2.5$ meter du Pont telescope \citep{SDSS_bowen1973} at Las Campanas Observatory using near-twin H-band spectrographs \citep{APOGEE_spec_wilson} as part of the fourth campaign of SDSS \citep[SDSS-IV;][]{blanton2017}. The APOGEE data reduction, Spectral Parameters,
and Chemical Abundances Pipeline \citep[ASPCAP;][]{2015AJ....150..173N,2016AJ....151..144G,shetrone2015,vsmith_apogee_dr16_2021} provides stellar spectra and computes several stellar parameters including $T_\mathrm{eff}$, surface gravity ($\log g$), metallicity ($[Fe/H]$), and surface rotational velocity ($v\sin i$). We specifically use propriety parameters from APOGEE Data Release 16 plus \citep[DR16]{Jonsson_2020,APOGEE_DR16,vsmith_apogee_dr16_2021} for stars in the TESS SCVZ and public data from the APOGEE DR16 standard release for stars in the \textit{Kepler} field. APOGEE DR16 plus includes observations up to March 2020 (MJD=58932) and APOGEE DR16 includes observations up to August 2018 (MJD=58358).

We augment our TESS-APOGEE SCVZ sample with stars observed by APOGEE and the \textit{Kepler} Space Telescope. We specifically select stars in the APOGEE-\textit{Kepler} sample that have values for rotation periods measured by \citet{nielsen_2013}, \citet{reinhold_2013}, or \citet{2014ApJS..211...24M} or are listed as nondetections in \citet{2014ApJS..211...24M}. For stars with rotation periods from more than one of these samples, we adopt the rotation period from \citet[][]{2014ApJS..211...24M}. When compared to the other studies via a hare-and-hounds exercise \citep[][]{2015MNRAS.450.3211A}, the \citet[][]{2014ApJS..211...24M} rotation periods were found to be more reliable than those found by \citet{nielsen_2013} and \citet{reinhold_2013}. We note that although other studies were found to have more reliable rotation periods than \citet[][]{2014ApJS..211...24M}, \citet{nielsen_2013}, and \citet{reinhold_2013} by \citet{2015MNRAS.450.3211A} \citep[e.g.][]{garcia_2014}, their samples contain far fewer stars. We limit the APOGEE-\textit{Kepler} sample using the same $\log g$ and $T_\mathrm{eff}$ constraints we used on the APOGEE-TESS SCVZ sample, $3<\log g<5$ and $5600K<T_\mathrm{eff}<8000K$. Of the 23,000 stars observed in the APOGEE-\textit{Kepler} sample, our final \textit{Kepler} sample contains 2101 stars. The bottom panel in Figure \ref{apo_sample} shows a Kiel diagram of all stars in the APOGEE-\textit{Kepler} sample along with our final \textit{Kepler} sample.

We finally use stellar parallaxes from \textit{Gaia} data release 2 \citep[DR2]{2018A&A...616A...1G}, K-band magnitudes from 2MASS \citep[][]{Skrutskie_2006_2MASS}, and $T_\mathrm{eff}$ and $[Fe/H]$ from APOGEE to improve our estimates of the radii of stars in our final sample. This is done using the direct method of \texttt{isoclassify} \citep{Huber2017,Berger_2020a}. There were no parallax measurements for 68 stars in our sample, therefore we exclude them from our analysis. \texttt{isoclassify} is currently being adapted to analyze \textit{Gaia} Early Data Release 3 \citep[][]{Gaia_edr3} photometry. This update to \texttt{isoclassify} was not complete at the time of writing. 

\subsection{Processing TESS Lightcurves}
The TESS SPOC pipeline currently provides lightcurves generated via Simple Aperture Photometry (SAP) and Pre-search Data Conditioning SAP (PDCSAP) \citep{Jenkins_2016}. PDCSAP lightcurves use singular value decomposition to remove trends caused by the spacecraft. However, some of these trends are on timescales similar to the rotation signals we want to recover. Because of the possibility of suppressing rotation signals, we choose to use SAP lightcurves for our analysis. 

\begin{figure}[ht!]
    \centering
    \includegraphics[width=0.485\textwidth]{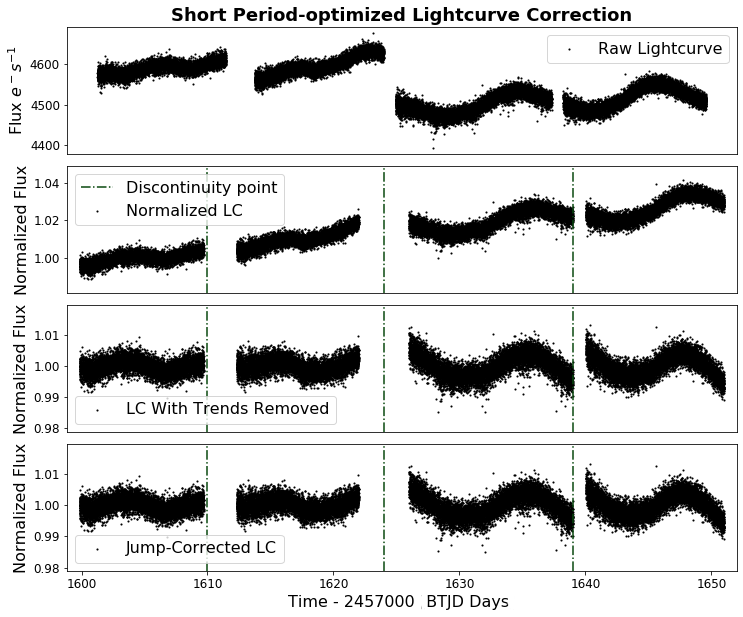}
    \caption{Creating short-period rotator-optimized lightcurves. The top panel shows the uncorrected SAP lightcurve for TIC 141482802. The second from top panel shows the sigma-clipped, normalized lightcurve with NaN values and values around discontinuous regions removed along with our identified discontinuity points. The third from top panel shows the flattened lightcurve with segment-length trends removed. The bottom panel shows the final lightcurve with all jumps corrected using the first segment as a reference. We use short-period rotator-optimized lightcurves in the remainder of our analyses.}
    \label{shortopt}
\end{figure}

To generate SAP lightcurves, the TESS pipeline sums the flux values of pixels within an aperture that optimizes the target star's signal-to-noise \citep{2017ksci.rept....6M}. However, this process retains significant instrument systematics caused by spacecraft pointing, safe modes, and scattered light. Additionally, our stars have been observed across at least nine sectors, and these systematics are observing sector-dependent. To correct and stitch together our lightcurves, we initially followed the methods performed in \citet{2011MNRAS.414L...6G} on the \textit{Kepler} sample. However, upon implementing this method on several simulated TESS lightcurves from \citet[][]{Claytor_2021}, we found that these methods preserved too many TESS-specific systematics for us to accurately measure rotation periods. In light of this, we devise our own method to correct for TESS systematics while preserving any possible rotation signals. 

We start by removing all NaN values from each lightcurve, sigma clipping any values greater than $4\sigma$ from the mean in each individual sector, and removing any data points with a nonzero quality flag. We then clip the first and last two days of each observing sector where scattered light effects are most likely to be present. We also clip regions around TESS safe modes. We identify and clip regions around discontinuities and jumps that are found in all lightcurves in our sample. We then identify timestamps where we expect the lightcurve to be discontinuous (e.g. at the start and end of each sector and at safe mode times), and median-normalize the flux in each part of the lightcurve. Portions of the lightcurve between discontinuities and the start and end of observing sectors are hereafter referred to as "segments".

\begin{figure}
    \centering
    \includegraphics[width=0.485\textwidth]{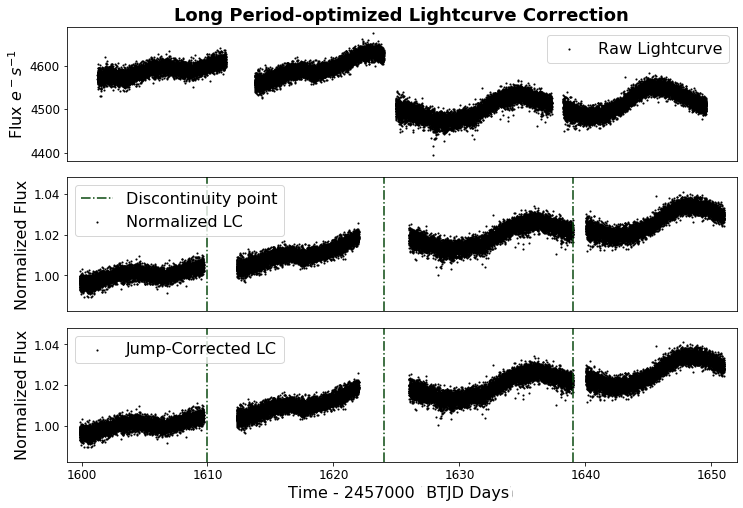}
    \caption{Creating long period rotator-optimized lightcurves. The top panel shows the uncorrected SAP lightcurve for TIC 141482802. The second from top panel shows the sigma-clipped, normalized lightcurve with NaN values and values around discontinuous regions removed along with our identified discontinuity points. The bottom panel shows the final lightcurve with all jumps corrected using each segment as the reference for the next segment. This method preserves systematic trends that are inconsistent with expectations from $v\sin i$ more than the short rotation period-optimized method, as seen by the $\sim50$ day linear increase in brightness. Because of this, we do not use long period rotator-optimized lightcurves in the remainder of our analyses.}
    \label{longopt}
\end{figure}

To correct discontinuous regions and long-term trends, we take a two-tiered approach optimized for the range of rotation periods we expect to find. We devise a method for stars with rotation periods less than one observing sector (hereafter referred to as short-period rotators) and for stars with rotation periods greater than one observing sector (hereafter referred to as long-period rotators).

For creating lightcurves optimized for short-period rotators, we do a linear fit to each segment. We then subtract this line and re-normalize each segment so the median value is 1. This yields flattened lightcurve segments that only differ by a scalar factor. We then compute the median flux of the first observed sector and use that as a reference. We correct the rest of the lightcurve using the difference between each segment's median flux and the reference flux. The steps for creating lightcurves optimized for short-period rotators are shown in Figure \ref{shortopt}.

\begin{figure*}[ht]
    \centering
    \includegraphics[width=0.95\textwidth]{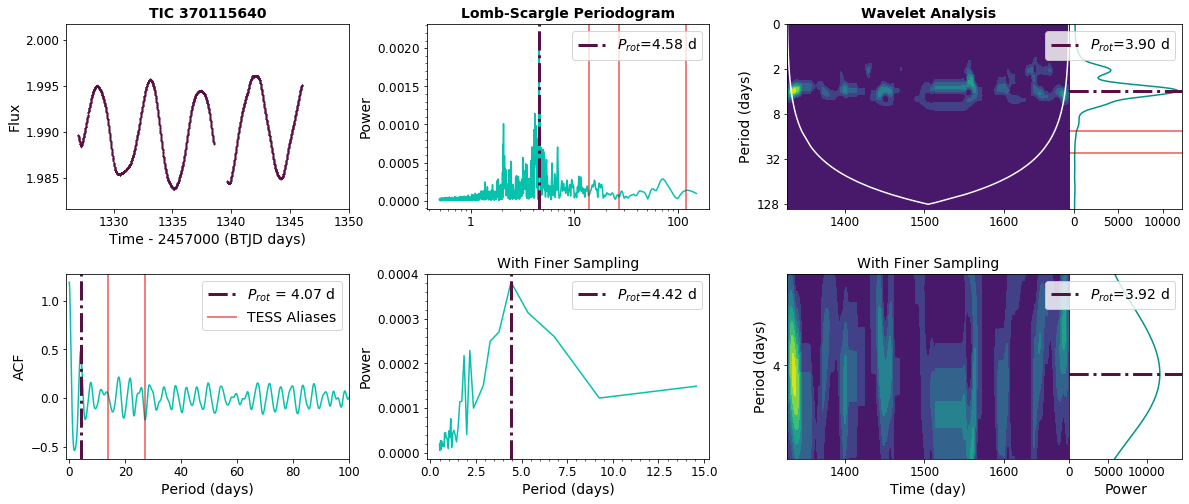}
    \caption{All of our rotation period measurement methods run on TIC 370115640. The top left panel shows a 30 day segment of the full lightcurve. The bottom left panel shows the autocorrelation function and its measured rotation period. The middle panels show the full Lomb-Scargle periodogram (top) and the zoomed-in periodogram (bottom). The right panels show the full wavelet power spectrum and global wavelet power spectrum (top) and the zoomed-in wavelet power spectrum and global wavelet power spectrum (bottom). Red lines in all plots correspond to periods associated with signals from the TESS instrument.}
    \label{prot_measure_fig}
\end{figure*}

We utilize a different approach for creating lightcurves optimized for long period rotators. We do not subtract any trends from the existing lightcurve to preserve any rotation periods that are longer than one sector. Instead of using the first observed segment as a reference, the segment directly before a discontinuity is used as a reference for the segment directly after the discontinuity. We correct each segment of the lightcurve using the difference between the median fluxes of each segment. The steps for creating lightcurves optimized for long period rotators are shown in Figure \ref{longopt}. After correcting all jumps in both sets of lightcurves, we smooth the lightcurve using a boxcar kernel of 0.5 days to remove any high-frequency noise while also preserving any possible short period rotation signals. 

We run both the long period rotator and short-period rotator-optimized methods on our full sample of stars. After doing this, we run our rotation period measurement methods (see Section \ref{prot_measure_sec}) and test the validity of the recovered periods on both sets of lightcurves (see Section \ref{prot_valid}). From visual inspection comparing both sets of lightcurves with the rotation periods we recover, we find that the long period rotator-optimized method still retains substantial systematic signals that both prevent the recovery of rotation signals longer than 13.7 days and suppress real rotation signals shorter than 13.7 days. Therefore our long period rotator-optimized method is not a viable method for processing lightcurves for stars observed in multiple TESS observing sectors. Because of this, we only use the short-period rotator-optimized lightcurves in this paper. 

\subsection{Measuring Rotation Periods in TESS}\label{prot_measure_sec}

After generating short-period rotator-optimized lightcurves for our full sample of stars, we compute rotation periods using a combination of Lomb-Scargle periodograms, wavelet analysis, and autocorrelation function (ACF) analysis. We apply these methods to all the stars in our sample. Previous work has shown that a combination of these methods optimizes the recovery of correct stellar rotation periods in injection and recovery exercises \citep{2015MNRAS.450.3211A,2017A&A...605A.111C}.

The Lomb-Scargle periodogram uses Fourier transforms to recover any periodic trends found in a dataset and their associated periods and shows which recovered periods have the strongest power using methods optimized for unevenly time-sampled data \citep{Scargle_1982ApJ...263..835S,reinhold_2013}. We utilize the Lomb-Scargle periodogram method within the lightcurve processing package \texttt{lightkurve} \citep{Lightkurve}. We ignore peaks in the periodogram that are within 10\% of periods related to the orbital period and observing cadence of the TESS instrument and their aliases at 13.7, 27.4, 90, and 120 days. We then recompute the periodogram to search the region 10 days around the highest peak in the initial periodogram with finer sampling. Our final rotation period from Lomb-Scargle periodograms is defined as the highest peak in the recomputed periodogram that does not correspond to a TESS alias. An example of our computed periodograms for TIC 370115640 is shown in the middle panel of Figure \ref{prot_measure_fig}.

\begin{deluxetable*}{ccccccccc}[ht!]
    \tablecaption{Properties of stars with recovered rotation periods, potential binaries, pulsating stars, and stars with no detected variability in our TESS and \textit{Kepker} samples. $T_\mathrm{eff}$, $\log g$, $v\sin i$, metallicity, and $\alpha$ abundance are from APOGEE spectra. Stellar radii and masses are estimated using \texttt{isoclassify} \citep[similar to]{Berger_2020a}. This table is available in its entirety in machine-readable format. A portion is shown here for guidance regarding its form and content.\label{starp_tbl}}
    \tablehead{\colhead{ID} & \colhead{Flag} & \colhead{$T_\mathrm{eff}$ (K)} & \colhead{$\log g$ (dex)} & \colhead{$v\sin i$ ($km s^{-1}$)} & \colhead{[$Fe/H$] (dex)} & \colhead{[$\alpha/M$] (dex)} & \colhead{$R$ ($R_{\odot}$)} & \colhead{$M$ ($M_{\odot}$)}}
    \startdata
        TIC 30631031 & Rotator & $7136\pm183$ & $3.89\pm0.07$ & 58.09 & $-0.02\pm0.03$ & $0.12\pm0.03$ & $2.21^{+0.05}_{-0.05}$ & $1.63^{+0.05}_{-0.05}$ \\
        TIC 55480220 & RUWE Binary & $6947\pm170$ & $3.98\pm0.07$ & $13.49$ & $0.13\pm0.02$ & $0.03\pm0.01$ & $1.98^{+0.05}_{-0.04}$ & $1.58^{+0.05}_{-0.05}$ \\
        TIC 150393064 & Spectroscopic Binary & $5868\pm121$ & $4.47\pm0.08$ & $20.06$ & $-0.02\pm0.01$ & $-0.04\pm0.01$ & $1.32^{+0.07}_{-0.07}$ & $1.11^{+0.05}_{-0.06}$ \\
        TIC 272127517 & Pulsator & $6774\pm163$ & $3.95\pm0.07$ & $80.05$ & $0.05\pm0.02$ & $0.04\pm0.02$ & $1.98^{+0.04}_{-0.04}$ & $1.50^{+0.04}_{-0.05}$ \\
    \enddata
\end{deluxetable*}

The Lomb-Scargle periodogram, however, falls short in that it is optimized for periodic phenomena. Often, a rotation signal will be a mixture of periodic and aperiodic signals, as starspots are not permanent features on a star's surface. Wavelet analysis overcomes perturbations in the lightcurve caused by the evolution of starspots by determining the periodic signals present in a slice of timeseries data and stepping through the rest of the time series to test the strength of those periods \citep{1998BAMS...79...61T}. We use the Morlet wavelet transform in the SciPy library \citep{2020SciPy-NMeth} with the power spectral density correction of \citet{Liu_2007} to perform these measurements. We first compute the wavelet power spectrum and global wavelet power spectrum (GWPS), which is equal to the wavelet power spectrum summed over the full timeseries, with coarse spacing in frequency. As with the Lomb-Scargle periodogram, we ignore any peaks in the wavelet power spectrum and GWPS that correspond to TESS instrument signals. We then recompute the wavelet power spectrum and the GWPS with finer sampling in the region within 10 days of the highest peak in the GWPS. Our final rotation period from wavelet analysis is defined as the highest peak in the zoomed-in GWPS that doesn't correspond to a TESS alias. An example of our computed wavelet power spectra and GWPS for TIC 370115640 is shown in the right panel of Figure \ref{prot_measure_fig}.

ACF analysis searches for periodic signals by correlating time series data with itself \citep{mcquillan_2013MNRAS.432.1203M}. This allows for confirmation of signals found through the Lomb-Scargle periodogram and wavelet analysis while also searching for signals that are not perfectly periodic, sinusoidally shaped, or present throughout the full dataset. We use the methods outlined in \citet{mcquillan_2013MNRAS.432.1203M} and the Python package \texttt{starspot} \citep[][]{starspot} to compute rotation periods with ACF analysis. As with the Lomb-Scargle periodogram and wavelet analysis, we ignore any peaks in the ACF that correspond to TESS instrument signals and their aliases. Our final rotation period from ACF analysis is defined as the first peak that does not correspond to a TESS alias or the second peak in the ACF if the second peak is larger than the first peak \citep{mcquillan_2013MNRAS.432.1203M}. An example of our computed ACF for TIC 370115640 is shown in the bottom left panel of Figure \ref{prot_measure_fig}.

We then compute the total uncertainty on our measured rotation periods. The statistical uncertainty for each of our methods is defined as the full-width at half maximum of the peak corresponding to the rotation period in the recomputed Lomb-Scargle periodogram, recomputed GWPS, and ACF. Additionally, \citet{2014ApJ...780..159E} argue that one should adopt a limit on the precision of any measured rotation period of 10\% due to stellar differential rotation. We propagate the statistical uncertainty from each of our rotation period measurement methods to obtain the final uncertainty on the photometric rotation period for each star. If that uncertainty is less than the assumed systematic floor from \citet{2014ApJ...780..159E}, we adopt 10\% as the final uncertainty. We list the properties of all stars in our sample, including those with detected variability and no detected variability, in Table \ref{starp_tbl}.

\subsection{Modeling}
Recent observations reveal that angular momentum loss is likely more complicated for stars unlike the Sun \citep[e.g.][]{2013ApJ...776...67V,kurtz_2014MNRAS.444..102K,Douglas_2017,2018ApJ...868..150T,curtis_2019ApJ...879...49C,Curtis_2020}. A recent parameterization, was devised by Pinsonneault, Matt, and MacGregor and was derived from magnetohydrodynamic simulations in \citet{2012ApJ...754L..26M} coupled to mass loss and magnetic field scalings \citep[see][]{2013ApJ...776...67V}. This law includes a stronger dependence on stellar radius and an explicit dependence on a star's convective overturn timescale, and is better able to match observed trends in hot ($>6000\,K$) and more evolved stars \citep{2013ApJ...776...67V,2018ApJ...868..150T}. Discussions of this law can be found in \citet{2012ApJ...754L..26M}, \citet{2013ApJ...776...67V}, and \citet{Amard_2020}. 

To analyze how well our measured rotation periods follow model predictions and to determine additional stellar parameters, we use the Python package \texttt{kiauhoku} \citep{2020ApJ...888...43C}. \texttt{kiauhoku} uses model grids computed using the Yale Rotating stellar Evolution Code \citep[YREC]{Pinsonneault_1989_yrec,2008Ap&SS.316...31D,tayar_2020}. A summary of the input physics used in the YREC models can be found in \citet{2013ApJ...776...67V} and \citet{2020ApJ...888...43C}. We specifically use model grids that compute a star's rotational evolution using the angular momentum loss law described \citet{2012ApJ...754L..26M,2013ApJ...776...67V}. \texttt{kiauhoku} has the ability to interpolate between model grids computed at multiple masses and metallicities to provide parameters of interest (e.g. $P_\mathrm{rot}$, radius, mass, and zero-age main sequence $T_\mathrm{eff}$) given observed values for $T_\mathrm{eff}$, $\log g$, $[Fe/H]$, and $[\alpha/M]$ from APOGEE.

\section{Rotation Period Validation}\label{prot_valid}

After collecting our final samples in \textit{Kepler} and TESS and running the rotation period measurement methods described in Section \ref{prot_measure_sec} on our TESS sample, we validate the rotation periods in both samples. 

\subsection{Pulsators in TESS}

\begin{figure}[ht]
    \centering
    \includegraphics[width=0.485\textwidth]{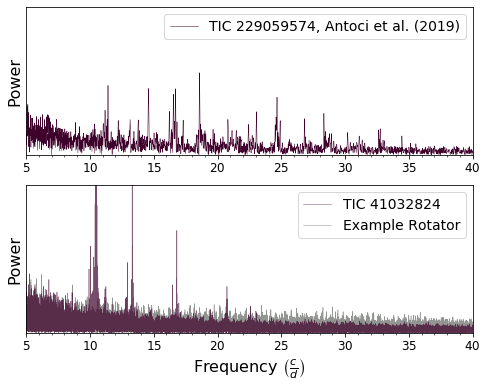}
    \caption{Identifying pulsators in our TESS sample. We show the Lomb-scargle periodograms in frequency space for a $\delta$ Scuti pulsator in TESS identified by \citet{Antoci_2019_puls} (top) and a candidate pulsator identified in this work (bottom, purple). We additionally plot an example rotator periodogram (grey) for TIC 370115640 in the bottom panel. We observe that the spacings between peaks in the periodograms are the same for the $\delta$ Scuti pulsator identified by \citet[][]{Antoci_2019_puls} and our candidate pulsator, indicating that the star in our sample is a likely pulsator.}
    \label{puls_id}
\end{figure}

Previous work has shown that pulsational and rotational modulation has been detected in lightcurves of stars in the $T_\mathrm{eff}$ range probed by our sample 
\citep[][]{2019MNRAS.485.2380M,Antoci_2019_puls}. We visually inspect all TESS lightcurves to determine whether the period found by our pipeline is more likely to be caused by pulsations than rotation. Pulsating stars are distinguished from rotators by multiple, regularly-spaced peaks in their Lomb-Scargle periodograms. Common pulsating contaminants in the region of the Kiel diagram occupied by our sample are $\delta$ Scuti and $\gamma$ Doradus type pulsators. $\delta$ Scuti type pulsators display equally spaced peaks in frequency from $5 d^{-1}$ to $50 d^{-1}$ \citep[see e.g.][]{2019MNRAS.485.2380M}, while $\gamma$ Doradus type pulsators display peaks that are equally spaced in period with frequencies below $5 d^{-1}$ (see Figure \ref{puls_id} for an example of a $\delta$ Scuti type pulsator). We find that our rotation period measurement pipeline incorrectly identifies these potential pulsation periods as rotation periods. We remove these stars from our final rotation sample and flag them for future users.

\begin{figure}[ht!]
    \centering
    \includegraphics[width=0.48\textwidth]{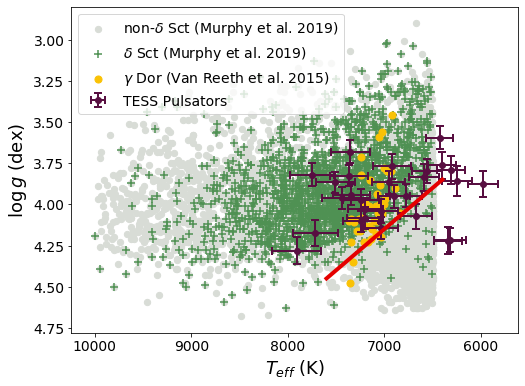}
    \caption{Kiel diagram comparing the pulsators identified in our TESS sample (purple) to the $\delta$ Scuti pulsators identified by \citet{2019MNRAS.485.2380M} (green) in their search of stars in the Kepler field (gray). We also show $\gamma$ Doradus pulsators identified by \citet{2015ApJS..218...27V} (orange) in the \textit{Kepler} sample. The pulsators in our sample follow the distribution of pulsators in the \textit{Kepler} sample and lie near the $\delta$ Scuti instability strip (red line) identified in \textit{Kepler} by \citet{2019MNRAS.485.2380M}, thus supporting our identification of them as pulsators.}
    \label{pulsators}
\end{figure}

Our visual inspection yields 28 potential pulsating stars in our TESS sample, which have been identified with the flag "Pulsator" in Table \ref{starp_tbl}. Figure \ref{pulsators} shows a Kiel diagram of our identified TESS pulsators along with $\delta$ Scuti and $\gamma$ Doradus pulsators found in the \textit{Kepler} sample by \citet{2019MNRAS.485.2380M} and \citet{2015ApJS..218...27V}. Our identified TESS pulsators clearly follow the $\delta$ Scuti instability strip identified in the \textit{Kepler} sample \citep{2019MNRAS.485.2380M}. Additionally, this initial investigation suggests that several of our pulsators also populate the $\delta$ Scuti - $\gamma$ Doradus hybrid regime, where stars are particularly useful testbeds for stellar physics \citep{2015ApJS..218...27V}. However, we leave confirmation and direct characterization of the pulsation modes of these stars to future, more targeted studies. 

\begin{figure}[ht!]
    \centering
    \includegraphics[width=0.48\textwidth]{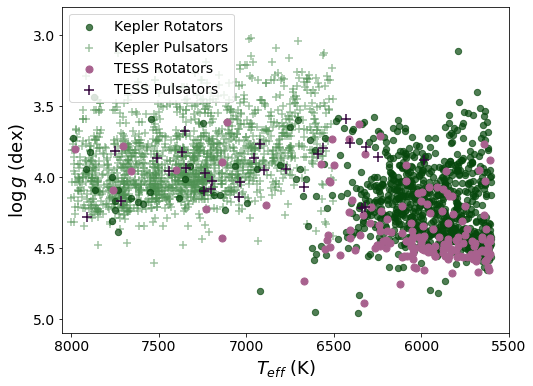}
    \caption{Kiel diagram comparing the pulsators and rotators identified in our TESS sample (purple) to the pulsators and rotators identified by \citet{2019MNRAS.485.2380M}, \citet{2015ApJS..218...27V}, \citet{nielsen_2013}, \citet{reinhold_2013}, and \citet{2014ApJS..211...24M} (green) in the \textit{Kepler} sample. The pulsators and rotators in both samples overlap in $T_\mathrm{eff}$ and$\log g$.}
    \label{kep_tess_puls}
\end{figure}

We compare our identified pulsators and rotators in TESS to those identified in \textit{Kepler} in Figure \ref{kep_tess_puls}. We find that pulsators and rotators overlap in the region between $6500\,K\,<T_\mathrm{eff}<7000\,K$. The \citet{2014ApJS..211...24M} study applied a cut at $T_\mathrm{eff} = 6500\,K$ when selecting their sample to avoid any pulsators when recovering rotation periods. However, previous studies \citep[e.g.][]{Antoci_2019_puls} and our results in TESS show that pulsators exist at effective temperatures less than $6500\,K$ and rotational variables with apparent starspot modulation exist at effective temperatures greater than $6500\,K$.

\subsection{Binaries in TESS and \textit{Kepler}}\label{sec_binaries}

We use the re-normalized unit-weight error (RUWE) statistic from \textit{Gaia} DR2 to identify any stars with likely wide (within 4", $\sim$1000 AU) binary companions in both the TESS and \textit{Kepler} samples \citep[see][for further discussion of the RUWE statistic]{Berger_2020a}. Following \citet{Berger_2020a}, we flag any stars with $RUWE>1.2$ as potential binaries. We find that 509 stars in our TESS sample (24.1\%) and 333 stars in our \textit{Kepler} sample (15.8\%) are flagged as potential wide binaries based on \textit{Gaia} RUWE. We identify these stars for future users with the flag "RUWE Binary" in Table \ref{starp_tbl}. 

We visually inspect APOGEE spectra for all stars in both our \textit{Kepler} and TESS samples to identify any likely double-lined spectroscopic (within 2", $\sim 500$ AU) binaries \citep[e.g.][]{fernandez2017_specbin,Kounkel2021arXiv210710860K}. We compare each spectrum to a model spectrum computed by ASPCAP, which assumes a spectrum for a single star. We flag stars that exhibit multiple spectral lines in regions where the ASPCAP model spectrum expects a single line as suspect. We find that 33 stars in our TESS sample and 20 stars in our \textit{Kepler} sample are likely spectroscopic binaries. We identify these stars for future users with the flag "Spectroscopic Binary" in Table \ref{starp_tbl}. Stars which were not identified as spectroscopic or RUWE binaries are referred to as "undetected" binaries in our following discussion since they may be single stars, but we cannot exclude the possibility of the undetected companion.

\subsection{Spectroscopic Validation}\label{spec_valid_sec}

We use $v\sin i$ from APOGEE and stellar radii estimated using \texttt{isoclassify} with \textit{Gaia} parallaxes and 2MASS photometry \citep[similar to][]{Berger_2018} to compute $P_\mathrm{rot}/\sin i$ for all stars in both our \textit{Kepler} and TESS samples. $P_\mathrm{rot}/\sin i$ is an upper limit on the rotation period we expect to observe from starspot modulation and is given by:

\begin{equation}
    P_\mathrm{rot}/\sin i = \frac{2\pi R_{Gaia}}{v\sin i}
\end{equation}

where $R_{Gaia}$ is the stellar radius constrained using \textit{Gaia} parallaxes. \citet{Simonian_2020} validated APOGEE $v\sin i$ values with rotation periods from \textit{Kepler} lightcurves and found a detection threshold of 10 $km s^{-1}$, where $v\sin i$ values less than 10 $km s^{-1}$ are considered suspect. \citet[][]{Tayar_2015} adopted a more generous threshold of 5 $km s^{-1}$ based off of comparisons between their measured $v\sin i$ values to previously published values. We adopt the same threshold as \citet[][]{Tayar_2015} and flag stars with ($v\sin i<$5 $km s^{-1}$) as suspect. Although we calculate $P_\mathrm{rot}/\sin i$ values for all stars, including those with unreliable $v\sin i$ values, we ignore these stars in our analysis of rotation periods derived with $v\sin i$.

We cross-validate our estimated stellar radii computed with the direct method in \texttt{isoclassify} to model radii predicted by YREC models with \texttt{kiauhoku}. We find that $3\%$ of our \textit{Kepler} sample and $3\%$ of our TESS sample have \textit{Gaia} radii that differ from YREC radii by more than $1.3$ solar radii, which is the median radius of stars in our sample and the radius difference we expect for equal-mass binaries. We find that of these stars with inconsistent \textit{Gaia} and YREC radii, $21\%$ in our \textit{Kepler} sample are classified as RUWE or spectroscopic binaries and $51\%$ in our TESS sample are classified as RUWE or spectroscopic binaries (see \ref{sec_binaries} for further discussion of our binary classification techniques). Given that an estimated $40\%$ of astrometric binaries in \textit{Gaia} are expected to not be detected and \textit{Gaia} is unable to detect binaries with angular separations below 20 milliarcseconds \citep[][]{2018A&A...616A...1G}, we attribute inconsistent \textit{Gaia} and YREC radii to undetected binaries. We additionally find that these inconsistent stars do not lie on the apparent main sequence when plotted on a \textit{Gaia} radius v. APOGEE $T_\mathrm{eff}$ diagram, supporting our suspicion that these are likely undetected binaries in \textit{Gaia}.  

Given our confidence in our APOGEE $v\sin i$ and \textit{Gaia} radii values, we validate our photometric rotation periods from TESS and \textit{Kepler} with $P_\mathrm{rot}/\sin i$.  We consider any photometric rotation periods longer than the expected maximum of  $P_\mathrm{rot}/\sin i$ as suspect, as there exists no inclination $i$ where the photometric rotation period can be greater than $P_\mathrm{rot}/\sin i$. Additionally, any rotation periods where the implied inclination angle is less than $20$ degrees are also flagged, as starspot modulation is typically not possible to detect in the lightcurves of these stars \citep{santos_2017}. We do this for both TESS rotation periods and our adopted \textit{Kepler} rotation periods computed by \citet{nielsen_2013}, \citet{reinhold_2013}, or \citet{2014ApJS..211...24M}. 

\begin{figure*}[ht]
    \centering
    \includegraphics[width=0.9\textwidth]{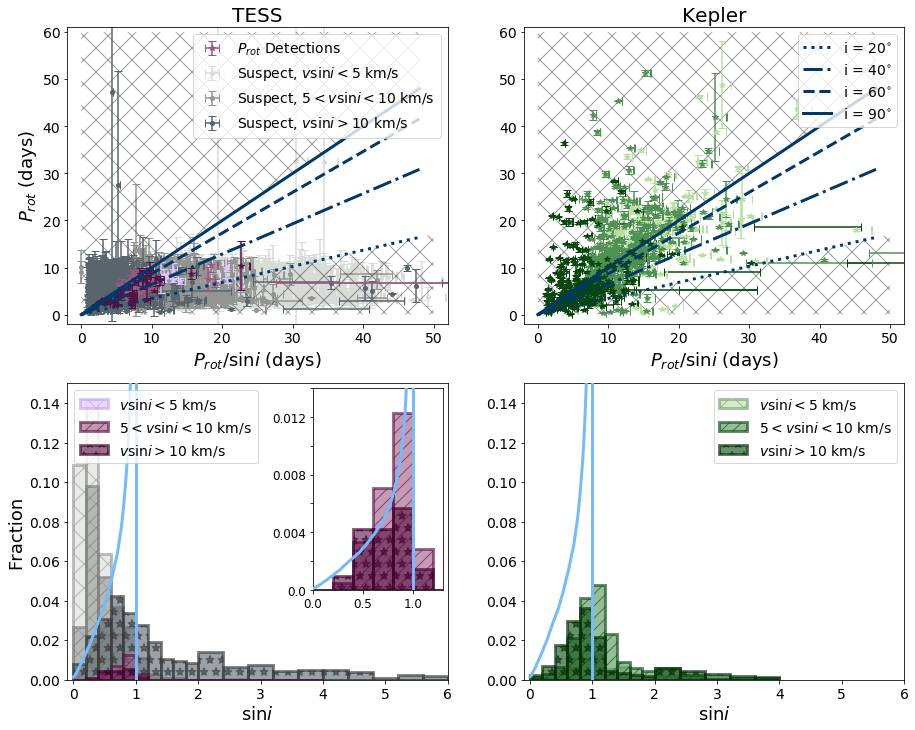}
    \caption{Validation of photometric rotation periods with $v\sin i$ from APOGEE and stellar radii estimated using \textit{Gaia}. The top panel shows the comparison between $P_\mathrm{rot}$ and $P_\mathrm{rot}/\sin i$ for our full TESS sample (left) and our \textit{Kepler} detection sample (right), including stars flagged as RUWE binaries. The different lines on both panels indicate where a star is expected to fall for a given inclination angle $i$, as shown in the right legend. The hatched regions indicate where $P_\mathrm{rot}$ and $P_\mathrm{rot}/\sin i$ are inconsistent (i.e. no real star can live in these regions). Stars flagged as suspect detections in our TESS sample are shown in grey, while stars in our TESS (\textit{Kepler}) sample with reasonable detections are shown in purple (green). The shade of points on the top panels indicate the trustworthiness of $v\sin i$ values used to calculate $P_\mathrm{rot}/\sin i$; the lightest color indicates that $v\sin i$ is far below the APOGEE detection threshold of $10\,km s^{-1}$, while the darkest color is above the APOGEE detection threshold. The bottom panel shows the distribution of $\sin i$ for the same TESS (left) and \textit{Kepler} (right) samples shown in the top panels, in the same color scheme as the top panels. The solid light blue line shows the expected distribution for a random sample of inclination angles. While the distribution of $\sin i$ values in our TESS detection sample does follow an expected distribution for a random sample of inclination angles, the \textit{Kepler} detection sample extends far above this distribution, as shown by the existence of many $\sin i$ values greater than 1, suggesting a problem with some of the \textit{Kepler} measurements.}
    \label{vsini_fig}
\end{figure*}

$P_\mathrm{rot}$ v. $P_\mathrm{rot}/\sin i$ and our inferred distributions of $\sin i$ for both our \textit{Kepler} and TESS samples are shown in Figure \ref{vsini_fig}. We find that for stars with reliable $v\sin i$ measurements, 536 stars in our TESS sample (41.3\%) and 957 stars in our \textit{Kepler} sample (66.9\%) have rotation periods that are inconsistent with $P_\mathrm{rot}/\sin i$ measured with $v\sin i$ from APOGEE spectra and stellar radii estimated using \textit{Gaia}, where stars are considered suspect detections when their inferred $\sin i$ values are greater than 1 or less than $\sin(20^{\circ})$. We additionally find that removing RUWE binaries does not significantly change our fraction of stars flagged as having reliable rotation periods, and therefore include them in Figure \ref{vsini_fig}. We identify stars with inconsistent $P_\mathrm{rot}$ and $P_\mathrm{rot}/\sin i$ for future users with the flag "Suspect Detection" in Table \ref{starp_tbl} and exclude them from our photometric rotation period analysis in Section \ref{lc_prot}.

The bottom right panel of Figure \ref{vsini_fig} shows that the distribution of $\sin i$ values in our \textit{Kepler} sample contains many stars with apparent $\sin i > 1$ (23.4\% of stars with reliable $v\sin i$ values in our \textit{Kepler} sample). The inferred \textit{Kepler} $\sin i$ distribution has many measured periods that cannot match the measured $v\sin i$ at any inclination angle, even after removing stars with suspect $v\sin i$ values as shown by the differently shaded distributions in the bottom right panel of Figure \ref{vsini_fig}. Additionally, the distribution of \textit{Kepler} $\sin i$ values is strikingly different from the expected distribution for a random sample of inclination angles (blue line in bottom right panel of Figure \ref{vsini_fig}). While we cannot conclusively identify the source of this inconsistency, previous work has suggested that many incorrect measurements of rotation periods in \textit{Kepler} lightcurves are caused by detections of rotation period aliases at twice the real rotation period \citep[][]{simonian_2019}. This hypothesis is additionally consistent with previous studies which found that 20-30\% of stars in the \textit{Kepler} sample will yield incorrect rotation periods even if they display starspot modulation in their lightcurves \citep{2015MNRAS.450.3211A,santos_2019}. 

\section{Photometric Rotation Periods}\label{lc_prot}

\subsection{Rotation in TESS}\label{tess_rot}

\begin{deluxetable*}{ccccccc}[ht!]
    \tablecaption{Rotation periods for our TESS sample measured with Lomb-Scargle periodograms, wavelet analysis, and autocorrelation function analysis, the mean photometric rotation period, and the rotation period derived from APOGEE spectra and \textit{Gaia} radii.\label{prot_tbl}}
    \tablehead{\colhead{ID} & \colhead{Binary Flag} & \colhead{Lomb-Scargle (d)} & \colhead{Wavelet (d)} & \colhead{ACF (d)} & \colhead{$P_\mathrm{rot}$ (d)} & \colhead{$P_\mathrm{rot}/\sin i$ (d)}}
    \startdata   
        TIC 30631031 & Single Star & $2.11\pm0.21$ & $1.93\pm0.19$ & $1.97\pm0.19$ & $1.99\pm0.20$ & $1.93^{+0.04}_{-0.04}$\\
        TIC 41591638 & Spectroscopic Binary & $7.31\pm0.73$ & $6.11\pm0.61$ & $10.23\pm1.02$ & $7.42\pm1.51$ & \nodata\\
        TIC 231072603 & RUWE Binary & $3.70\pm0.37$ & $3.53\pm0.35$ & $3.30\pm0.33$ & $3.52\pm0.35$ & $4.38^{+0.13}_{-0.12}$\\
    \enddata 
\end{deluxetable*}


Our final detection sample in TESS includes 169 rotation periods, which represents $7.9\%$ of our original TESS sample of 2115 stars. This sample includes stars that have reliable $P_\mathrm{rot}/\sin i$ measurements, where $v\sin i$ is greater than 5 $km s^{-1}$. $P_\mathrm{rot}$ can therefore be vetted to be consistent with $P_\mathrm{rot}/\sin i$. Of these stars with confidently recovered rotation periods, 20 are flagged as RUWE binaries and 1 is flagged as a spectroscopic binary. We list our measured rotation periods in Table \ref{prot_tbl}.

\begin{figure}[ht!]
    \centering
    \includegraphics[width=0.48\textwidth]{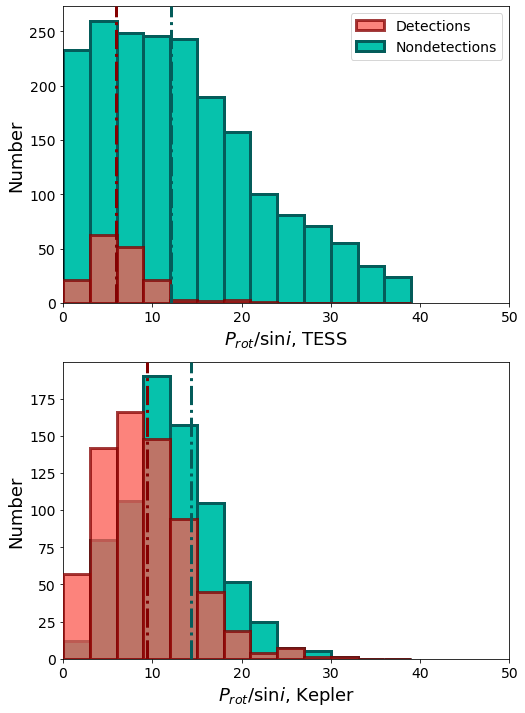}
    \caption{Distribution of spectroscopic rotation periods in our TESS (top) and \textit{Kepler} (bottom) samples for both detections (orange) and nondetections (turquoise). Here, nondetections refer to stars which did not exhibit spot modulation in their lightcurves. Additionally, these distributions only include stars with reliable $v\sin i$ values (i.e. $v\sin i > 5$ $km s^{-1}$). The dashed lines correspond to the median of each distribution. TESS rotation periods are clearly biased to those less than 13.7 days. The \textit{Kepler} sample also shows a bias to shorter rotation periods, but reaches rotation periods out to 30 days.}
    \label{det_lim}
\end{figure}

We posit that our ability to detect rotation periods greater than $13.7$ days is suppressed by TESS systematics, thus biasing our measured rotation periods to more rapidly rotating stars. This detection limit is shown in Figure \ref{det_lim}, where we compare the distribution of rotation periods detected in TESS to those detected in \textit{Kepler}, where rotation periods up to tens of days were measured. Although other attempts have been made to measure rotation periods for stars that span multiple TESS observing sectors \citep[e.g.][]{Lu_astrea_2020,Martins_2020_tessrot,Claytor_2021}, up to this point, only studies using machine learning-based rotation period extraction methods, i.e. neural networks, have successfully recovered rotation periods longer than $13.7$ days \citep[][]{Lu_astrea_2020,Claytor_2021}.

\subsection{Comparing TESS and Kepler}
\begin{figure*}[ht!]
    \centering
    \includegraphics[width=0.9\textwidth]{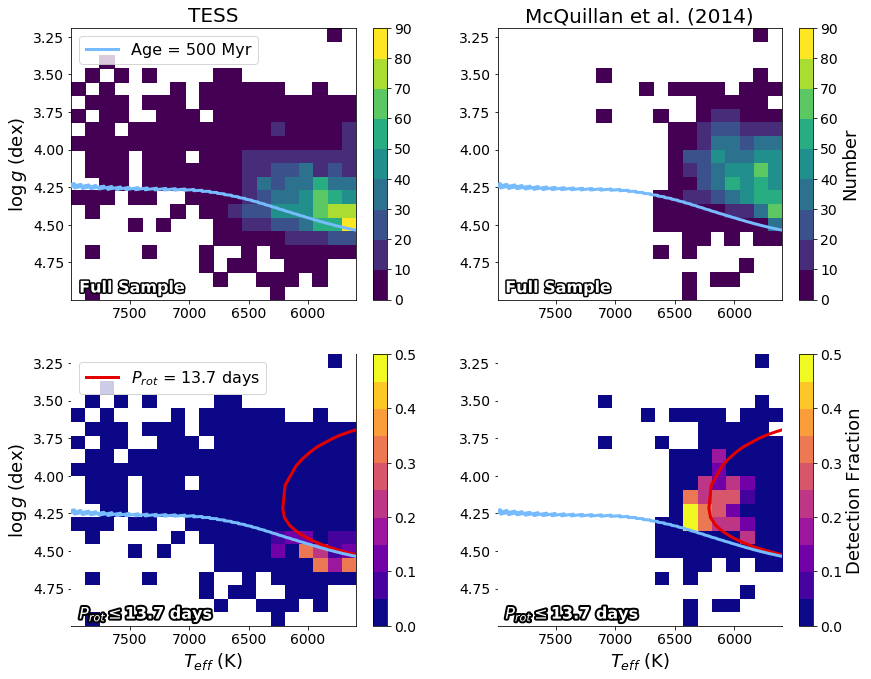}
    \caption{Kiel diagram of stars with detected rotation periods in TESS (left) and the \citet[][]{2014ApJS..211...24M} \textit{Kepler} sample (right). The top panels show the number of stars in the full (combined detection and nondetection) samples in each $T_\mathrm{eff}$ - $\log g$ bin, while the bottom panels show the detection fraction of photometric rotation periods in each bin. The detection fraction distribution for the \textit{Kepler} sample (bottom right) is calculated only for stars whose detected rotation periods are less than $13.7$ days. The lines correspond to a constant age of $500$ Myr (blue) and a constant rotation period of $13.7$ days (red) as determined by \texttt{kiauhoku} from YREC models at solar metallicity.}
    \label{det_frac}
\end{figure*}


Prior to analyzing distributions of stellar parameters as a function of rotation in our TESS and \textit{Kepler} samples, we first examine the differences present in these two populations. We compare the detection fraction of stars across the Kiel diagram in the \citet{2014ApJS..211...24M} sample and our TESS sample. We restrict our comparison in the \citet{2014ApJS..211...24M} sample to only include stars with detected rotation periods less than $13.7$ days to enable a more direct comparison to our biased TESS sample. Our comparison is shown in Figure \ref{det_frac}. The top two panels show the number density of stars in our entire TESS sample and the entire \citet{2014ApJS..211...24M} sample. The bottom two panels show the distribution of detection fractions for stars in our TESS sample with detected photometric rotation periods (left) and stars in the \citet{2014ApJS..211...24M} sample with detected rotation periods less than $13.7$ days. We find that we are more likely to detect rotation periods for cooler, less evolved stars in TESS. This is consistent with \citet{2014ApJS..211...24M}, who observed that the amplitude of periodic variability ($R_{per}$) increased for cooler stars. 

We maximize our detection fraction for the coolest main sequence stars in our TESS sample ($T_\mathrm{eff}$ = $5600K - 5720K$, $\log g$ = $4.55 - 4.62$) at $32\%$ of stars with detected rotation periods. The \citet{2014ApJS..211...24M} sample contains 4 stars with detected rotation periods less than 13.7 days in this bin. To more accurately compare the detection fractions of our TESS sample and the \citet{2014ApJS..211...24M} sample, we take a wider range in $T_\mathrm{eff}$ and $\log g$ of $5600K - 5800K$ and $4.5 - 4.7$ dex. The detection fraction for TESS stars in these bins is 25\%, compared to the 40\% detection fraction seen in the \citet{2014ApJS..211...24M} sample, which includes measured rotation periods out to tens of days. This discrepancy in rotation period detection fractions is likely due to difficulties in measuring rotation periods longer than 13.7 days in TESS lightcurves, where any star in our TESS sample with a rotation period longer than 13.7 days is likely to have its rotation signal suppressed by TESS instrument signals. 

The distribution of detection fractions also differ substantially between our TESS sample and the \citet[][]{2014ApJS..211...24M} sample. We argue that these differences in rotation distributions between the \citet{2014ApJS..211...24M} sample and the TESS sample are likely caused by differences in stellar populations along the sightlines of \textit{Kepler} and TESS. The blue line in Figure \ref{det_frac} is a 500 Myr solar metallicity isochrone calculated by \texttt{kiauhoku}. The \textit{Kepler} sample includes few stars below the 500 Myr isochrone \citep[][]{Berger_2020a,Lu_2021_gyro}. This makes sense given the location of the \textit{Kepler} observing field above the galactic midplane, where few young stars reside. The existence of short-lived massive O and B stars \citep{Pedersen_2019,Bowman_2020} as well as the existence of stars identified with young ($<200$ Myr) clusters and moving groups \citep[e.g.][]{Gagne_2018_youngcluster} in the TESS SCVZ, suggests a young population is present in the direction of the TESS SCVZ that was not present in the \textit{Kepler} field. While we leave the creation of a full population model in the TESS SCVZ and analysis of age proxies (e.g. kinematics) to future studies, a rough comparison of our stars with existing rotation-age relationships \citep{2020ApJ...888...43C} suggests that this population likely covers a large age range, including a significant number of young stars. 

\begin{figure*}[ht!]
    \centering
    \includegraphics[width=0.9\textwidth]{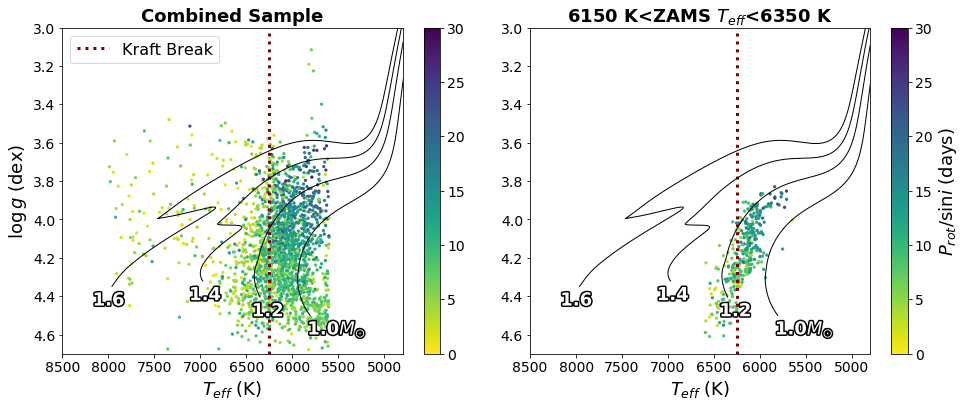}
        \caption{Rotation as a function of evolution in our combined TESS and \textit{Kepler} sample. When we show our entire sample (left), we recover the broad trend that more massive stars rotate more rapidly and that stars rotate faster with increasing $T_\mathrm{eff}$. However, when we include stars within a narrower range of zero-age main sequence $T_\mathrm{eff}$, we can see the declining rotation rate as stars evolve up the subgiant branch, expanding and losing angular momentum.}
    \label{hr_rotevol}
\end{figure*}

\subsection{Inconsistencies with Models}

We find inconsistencies between detected rotation periods in \textit{Kepler} and the predictions of models that include rotation. The red line in Figure \ref{det_frac} is the predicted location of stars with a rotation period of 13.7 days \citep[][]{2013ApJ...776...67V}. Under the conditions of this model, no periods shorter than this should be detected to the right of this line. The detection fraction in our TESS sample does drop significantly at around this line (bottom left panel of Figure \ref{det_frac}), suggesting consistency with the \citet[][]{2013ApJ...776...67V} prediction. The \textit{Kepler} sample (bottom right panel), however, is inconsistent with the \citet[][]{2013ApJ...776...67V} prediction. In fact, we see a significant overdensity in \textit{Kepler} short period rotators in this regime, comprising $\sim$70\% of stars with detected rotation periods less than $13.7$ days. Visual inspection of \textit{Kepler} target pixel files and our analysis of $v\sin i$ does not suggest an obvious abundance of background stars which could be responsible for this signal. The overdensity also does not coincide with the presence of low-amplitude rotation signals, where measurement errors are more likely to occur. Theoretical explanations for this overdensity of rapidly rotating slightly evolved stars are also unable to explain the data. We find that these stars are spinning too fast even when compared to models with weakened magnetic braking \citep[][]{2016Natur.529..181V}. Additionally, the metallicity distribution of these anomalously fast rotating stars is consistent with the general sample, so explanations relying on metallicity-dependent rotational evolution are unlikely \citep[][]{Amard_2020}.

We therefore posit that the overdensity at rotation periods longer than $13.7$ days in the \citet{2014ApJS..211...24M} sample is most likely due to unidentified binaries in the \textit{Kepler} field. As stated in Section \ref{spec_valid_sec}, an estimated 40\% of astrometric binaries and binaries with separations below 20 milliarcseconds will not be detected in \textit{Gaia} DR2 \citep[][]{2018A&A...616A...1G}. Although upon removing RUWE and spectroscopic binaries from the \citet{2014ApJS..211...24M} sample, we find no significant change in the detection fractions we compute, given the high fraction of undetected binaries in \textit{Gaia}, we still suspect that this overdensity is most likely due to unidentified binaries. This conclusion is consistent with previous work that attributed anomalously rapid rotation in \textit{Kepler} to undetected binary companions from comparisons with \textit{Gaia} \citep[][]{simonian_2019}.

\section{Probing Rotation with $P_{\mathrm{\lowercase{rot}}}/\sin \lowercase{i}$} 

Due to the substantial detection bias towards stars with rapid rotation in our TESS sample, any distributions we derive from photometric rotation periods measured with TESS will be incomplete. Rotation periods derived from spectroscopic $v\sin i$, however, are distributed similarly in the TESS and \textit{Kepler} samples (see Figure \ref{det_lim}). We therefore use $P_\mathrm{rot}/\sin i$ for both our TESS and \textit{Kepler} samples for the remainder of our analysis. 

Although TESS and \textit{Kepler} probe different stellar populations, we expect the fundamentals of angular momentum acquisition and loss to be the same. We therefore combine our entire TESS and \textit{Kepler} samples into one larger sample for this section of our analysis. This combined sample includes all stars in our original \textit{Kepler} and TESS samples defined in Section \ref{data_obs} with $v\sin i$ values above a detection threshold of $5\,km s^{-1}$. 

\subsection{Rotation on the Kiel Diagram}\label{dist_rotevol}

In the regime probed by our combined \textit{Kepler} and TESS sample, we expect rotation to be a stronger function of $T_\mathrm{eff}$ than age. As described in Section \ref{intro}, hot stars ($T_\mathrm{eff}\,>$ 6250 K) do not experience substantial spin down on the main sequence, and therefore exhibit rapid rotation at the end of the main sequence. Cool stars ($T_\mathrm{eff}\,<$ 6250 K), however, are able to develop deep convective zones and thus sustain a magnetized wind and shed angular momentum on the main sequence.

We find that when we observe rotation across the Kiel diagram (see Figure \ref{hr_rotevol}) for our entire sample, rotation is indeed a strong function of $T_\mathrm{eff}$. At the Kraft Break, we find that the transition between slow ($P_\mathrm{rot} > 10$ days) and fast ($P_\mathrm{rot}<10$ days) rotators is apparent. We narrow the range of probed zero-age main sequence $T_\mathrm{eff}$ to include only stars within $100$ K of the Kraft Break. When we limit our sample to this range, we find that rotation becomes a weaker function of $T_\mathrm{eff}$ and a stronger function of evolutionary state. We additionally find that these rotation periods are inconsistent with models which include standard angular momentum loss \citep[][]{2013ApJ...776...67V} and weakened braking \citep[][]{2016Natur.529..181V}, with our stars, on average, rotating faster than predictions by both models. We suspect that the challenges with detecting slow rotational velocities may be the culprit (see Figure \ref{det_lim}). Future studies focused on testing model accuracy would benefit from samples which probe stars with a wide range of zero-age main sequence $T_\mathrm{eff}$ and from the detection of rotation periods beyond those probed by our sample. 

\subsection{Metallicity Dependence}\label{dist_metal}

\begin{figure}
    \centering
    \includegraphics[width=0.48\textwidth]{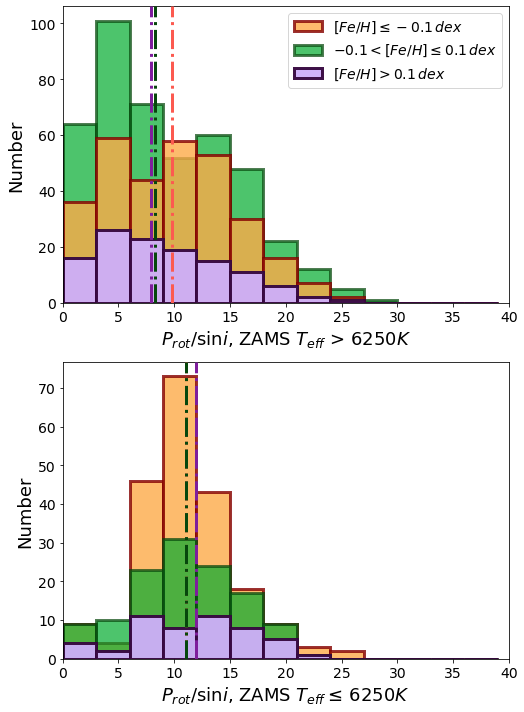}
    \caption{Distribution of rotation as a function of zero-age main sequence $T_\mathrm{eff}$ and metallicity. We divide our sample into a bin above the Kraft Break (top) and a bin below the Kraft Break (bottom), and divide our sample into low (orange), solar (green), and high (purple) metallicity bins. We find that neither zero-age main sequence $T_\mathrm{eff}$ bin exhibits a correlation between rotation and metallicity.}
    \label{psini_massfebin}
\end{figure}

\citet{Amard_2020} observed a correlation between rotation and metallicity in their sample of 4055 \textit{Kepler} stars between $0.85M_{\odot}$ and $1.3M_{\odot}$, with metal-rich stars rotating more slowly than metal-poor stars within this mass range. This result is presented as evidence for metallicity-dependent magnetic braking, and is consistent with trends between rotation and metallicity that are predicted for cool, old dwarfs \citep[][]{2013ApJ...776...67V,2020ApJ...888...43C,Simonian_2020}. The size of convection zones in these stars are sensitive to metallicity, where higher metallicity causes deeper convection zones at fixed mass \citep[][]{2013ApJ...776...67V}. Recently, the relation between stellar activity, convection zone depth, and metallicity has been observed in \textit{Kepler} by \citet[][]{See_2021}, who found that the amplitude of stellar variability exhibits a positive correlation with metallicity and the convective turnover timescale. 

Our sample, however, shows no correlation between rotation and metallicity. A Kolmogorov-Smirnov test supports that the metallicity-binned $P_\mathrm{rot}/\sin i$ distributions in Figure \ref{psini_massfebin} are drawn from the same distribution, indicating that our sample does not provide evidence for metallicity-dependent magnetic braking. 

This lack of a correlation between rotation and metallicity suggests that the rotational evolution of stars in our sample is not highly metallicity-dependent and suggests that our stars are experiencing weak magnetic braking on the main sequence. Although our sample includes a large number of stars below the Kraft break, where stars are expected to experience substantial spin-down, \citet[][]{Amard_2020} searches a cooler and lower-mass population of stars than us. When we limit our sample to stars cooler than $6000$K, we observe suggestions of a correlation between rotation and metallicity. However, these trends are not statistically significant. Further analysis of samples that specifically probe stars at a fixed mass and a wide range of metallicities is needed to determine the degree of correlation between metallicity and rotation. 

\subsection{Impact of Binarity}

\begin{figure}[ht!]
    \centering
    \includegraphics[width=0.48\textwidth]{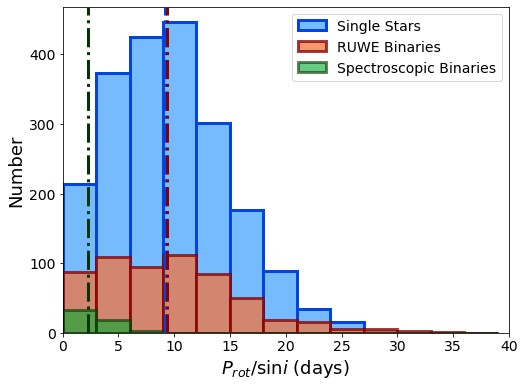}
    \caption{Distribution of spectroscopic rotation periods for single stars (blue), RUWE (wide) binaries (orange), and spectroscopic (close) binaries (green). The dashed lines correspond to the median of each distribution.}
    \label{hist_prot_binary}
\end{figure}

Binarity can impact a star's rotational history, from formation to tidal synchronization and mergers \citep[see e.g.][using \textit{Kepler}]{Douglas_2017,lurie_2017AJ....154..250L,fleming_2019}. We therefore use APOGEE spectra to identify double-lined close binaries (within $2"$, $\sim 500$ AU) and the RUWE statistic from \textit{Gaia DR2} to identify wide binaries (within $4"$, $\sim 1000$ AU) to determine if their rotation rates differ from a population of single stars. We show in Figure \ref{hist_prot_binary} that presumed single stars have a wide distribution of spectroscopic rotation periods ($P_\mathrm{rot}/\sin i$), out to 40 days, which is consistent with previous figures (see e.g., Section \ref{lc_prot}) and past studies with \textit{Kepler} \citep[see e.g.][]{nielsen_2013,reinhold_2013,2014ApJS..211...24M,Amard_2020}. Looking at spectroscopic (close) binaries, we observe a much more rapid median rotation rate of 2.3 days. Previous work suggests that rapid rotation in the \textit{Kepler} field is caused by a combination of rapidly rotating near-equal luminosity spectroscopic (close) binaries that have been spun up via tides and unresolved spectroscopic (close) binaries with velocity displacements on the order of APOGEE's resolution \citep{simonian_2019,Simonian_2020}. 

We additionally find that RUWE (wide) binaries have the same distribution in rotation as single stars at a $\approx 99\%$ significance, as supported by a Kolmogorov-Smirnov test \citep[see also][]{mamajek_2008,fleming_2019,godoy_rivera_2021}. By showing that single stars and the components of wide binaries have the same rotational distribution, Figure \ref{hist_prot_binary} validates the use of wide separation binaries as empirical constraints for the rotational evolution of single stars \citep[e.g.][]{mamajek_2008,godoy_rivera_2018,Janes_2017,fleming_2019,Otani_2021}.

\section{Summary \& Conclusions}

Using a sample of stars observed with TESS, rotation periods measured with \textit{Kepler}, spectroscopic parameters from APOGEE, and radii estimated using \textit{Gaia} parallaxes, we analyze rotation around the Kraft Break. With this sample, we show:

\begin{itemize}
    \item It is possible to detect rotation from starspot modulation in 2 minute cadence TESS SAP lightcurves using Lomb-Scargle, wavelet, and ACF analysis. However, TESS systematics suppress rotation periods longer than the TESS orbital period of $13.7$ days.
    \item The detection fraction for rotation periods in TESS is $7.6\%$, compared to the $20\%$ detection fraction seen in \textit{Kepler} by \citet[][]{Simonian_2020}. We suspect that this large discrepancy in detection fractions is primarily due to instrument systematics in TESS lightcurves.
    \item Spectroscopy can be sensitive to longer rotation periods that are suppressed by systematics in TESS lightcurves.
    \item We can identify other classes of stellar variability in TESS lightcurves, specifically pulsations (see Table \ref{starp_tbl}). These identified candidate pulsators show that there is a significant overlap in $T_\mathrm{eff}$ between rotating and pulsating stellar populations.
    \item Fundamental differences exist between the stellar populations probed by the \textit{Kepler} and TESS missions, with the TESS SCVZ field containing a significant number of young stars (Age < 500 Myr).  
    \item We recover broad trends between rotation and mass. From these trends, we find that tracing rotational evolution is challenging unless individual stars have been carefully characterized.
    \item Our sample does not appear to exhibit a correlation between rotation and metallicity; we argue that this is a result of the differences between the populations probed by our sample and the sample in \citet[][]{Amard_2020}. Our sample probes fewer cool low-mass stars and thus cannot probe metallicity-dependent spin down.
    \item A significant population of RUWE (wide) and spectroscopic (close) binaries are present in TESS SCVZ and \textit{Kepler}, comprising 24.6\% and 16.9\% of stars in these samples. Consistent with previous results, we find that spectroscopic (close) binaries rotate slightly faster than presumed single stars, whereas the rotation distribution of wide (RUWE) binaries covers the same range of rotation periods as single stars.
\end{itemize}

It is clear that stellar rotation is a complex physical process that has connections to our understanding of stellar interiors, activity, magnetism, and our own host star (the Sun). Past studies of rotational evolution have been successful in improving our understanding of rotation across stellar populations \citep[e.g.][]{2014ApJS..211...24M,Amard_2020,Simonian_2020}. The advent of comprehensive spectroscopic follow-up with APOGEE and photometric characterization with \textit{Gaia} have improved our ability to understand the multi-faceted nature of stellar rotation. Future spectroscopic surveys (e.g. SDSS-V and the Rubin Observatory) and future \textit{Gaia} releases \citep[e.g. EDR3;][]{Gaia_edr3} will enable precise characterization of stars across the entire sky. Therefore, further studies will be able to utilize a combination of photometric and spectroscopic parameters for a variety of stellar populations.

Strides are additionally being made in detecting long rotation periods from TESS lightcurves using more modern techniques like machine learning \citep[][]{Lu_astrea_2020,Claytor_2021}. Improved characterization of the TESS instrument as well as improved understanding of scattered light across all TESS observing sectors will improve our ability to detect rotation signals beyond tens of days. Additionally, future surveys would benefit from long baselines which allow studies to probe stars with a wide range of rotation periods. Detailed surveys (e.g. the European Space Agency PLAnetary Transits and Oscillations (PLATO) mission) and follow-up missions will improve our coverage of rotation across the entire sky, revealing the precise dependence of rotation on mass, metallicity, evolutionary state, and binarity. 

\acknowledgements 
E.A.A., J.N.T., and J.V.S. acknowledge support from NASA Award 80NSSC20K0056.
 
J.N.T. acknowledges that support for this work was provided by NASA through the NASA Hubble Fellowship grant No.51424 awarded by the Space Telescope Science Institute, which is operated by the Association of Universities for Research in Astronomy, Inc., for NASA, under contract NAS5-26555.

T.A.B. acknowledges support by a NASA FINESST award (80NSSC19K1424). A portion of T.A.B.’s research was supported by an appointment to the NASA Postdoctoral Program at the NASA Goddard Space Flight Center, administered by Universities Space Research Association and Oak Ridge Associated Universities under contract with NASA.

Support for this work was provided by NASA through Hubble Fellowship grant \#51386.01 awarded to R.L.B. by the Space Telescope Science Institute, which is operated by the Association of  Universities for Research in Astronomy, Inc., for NASA, under contract NAS 5-26555.

This paper includes data collected by the TESS mission, which are publicly available from the Mikulski Archive for Space Telescopes (MAST). Funding for the TESS mission is provided by NASA's Science Mission directorate.

Funding for the Sloan Digital Sky Survey IV has been provided by the Alfred P. Sloan Foundation, the U.S. Department of Energy Office of Science, and the Participating Institutions. SDSS acknowledges support and resources from the Center for High-Performance Computing at the University of Utah. The SDSS website is \url{www.sdss.org}.

SDSS is managed by the Astrophysical Research Consortium for the Participating Institutions of the SDSS Collaboration including the Brazilian Participation Group, the Carnegie Institution for Science, Carnegie Mellon University, the Chilean Participation Group, the French Participation Group, Harvard-Smithsonian Center for Astrophysics, Instituto de Astrof\'isica de Canarias, The Johns Hopkins University, Kavli Institute for the Physics and Mathematics of the Universe (IPMU) / University of Tokyo, the Korean Participation Group, Lawrence Berkeley National Laboratory, Leibniz Institut f\"ur Astrophysik Potsdam (AIP), Max-Planck-Institut f\"ur Astronomie (MPIA Heidelberg), Max-Planck-Institut f\"ur Astrophysik (MPA Garching), Max-Planck-Institut f\"ur Extraterrestrische Physik (MPE), National Astronomical Observatories of China, New Mexico State University, New York University, University of Notre Dame, Observat\'orio Nacional / MCTI, The Ohio State University, Pennsylvania State University, Shanghai Astronomical Observatory, United Kingdom Participation Group, Universidad Nacional Autónoma de M\'exico, University of Arizona, University of Colorado Boulder, University of Oxford, University of Portsmouth, University of Utah, University of Virginia, University of Washington, University of Wisconsin, Vanderbilt University, and Yale University.

This work has made use of data from the European Space Agency (ESA) mission \textit{Gaia} (\url{https://www.cosmos.esa.int/gaia}), processed by the \emph{Gaia} Data Processing and Analysis Consortium (DPAC, \url{https://www.cosmos.esa.int/web/gaia/dpac/consortium}). Funding for the DPAC has been provided by national institutions, in particular the institutions participating in the \emph{Gaia} Multilateral Agreement.

This publication makes use of data products from the Two Micron All Sky Survey, which is a joint project of the University of Massachusetts and the Infrared Processing and Analysis Center/California Institute of Technology, funded by the National Aeronautics and Space Administration and the National Science Foundation.

\software{Astropy \citep{Astropy,Astropy_2018}, isoclassify \citep{Huber2017,Berger_2020a}, Kiauhoku \citep{2020ApJ...888...43C},  Lightkurve \citep{Lightkurve}, Matplotlib \citep{Matplotlib}, NumPy \citep{numpy}, SciPy \citep{2020SciPy-NMeth}, and Starspot \citep[][]{starspot}.}

\facilities{Du Pont (APOGEE), Sloan (APOGEE), 2MASS,  \emph{Gaia}, TESS, \emph{Kepler}}

\bibliographystyle{aasjournal}
\bibliography{refs}

\end{document}